\documentclass[10pt]{iopart}
\usepackage{graphicx}
\usepackage{color}
\usepackage{siunitx}

\usepackage{iopams}  
\begin{document}

\title[SQUIDs on HgTe]{Phase-sensitive SQUIDs based on the 3D topological insulator HgTe}

\author{L. Maier, E. Bocquillon, M. Grimm, J.B. Oostinga, C. Ames, C. Gould, C. Br\"une, H. Buhmann, L.W. Molenkamp}

\address{Physikalisches Institut (EP3), Universit\"at W\"urzburg, Am Hubland, D-97074 W\"urzburg, Germany}
\ead{erwann.bocquillon@physik.uni-wuerzburg.de}
\vspace{10pt}

\begin{abstract}
Three-dimensional topological insulators represent a new class of materials in which transport is governed by Dirac surface states while the bulk remains insulating. Due to helical spin polarization of the surface states, the coupling of a 3D topological insulator to a nearby superconductor is expected to generate unconventional proximity induced $p$-wave superconductivity. We report here on the development and measurements of SQUIDs on the surface of strained HgTe, a 3D topological insulator, as a potential tool to investigate this effect.

\end{abstract}

%
%
%
%
\ioptwocol

\section{Introduction}

The prediction and discovery of topological insulators \cite{Kane2005,Bernevig2006,Konig2007,Hsieh2008,Xia2009,Bruene2011} have motivated tremendous experimental and theoretical efforts, aiming at understanding the peculiar properties of these materials. The electronic transport in such systems is mediated by conducting states with linear dispersion at the boundaries of the material, while the bulk remains insulating.

Remarkably, these surface states are predicted to exhibit helical spin polarization, with spin orthogonal to momentum. As a consequence, the pairing potential induced by a nearby conventional ($s$-wave) superconductor is expected to generate unconventional superconductivity, mixing $s$- and $p$- type characteristics \cite{Fu2008, Beenakker2013, Tkachov2013}. 

Here we report on the development and measurements of superconductor-topological insulator hybrid structures designed to investigate an expected topological superconductivity \cite{Maier2012, Veldhorst2012,Oostinga2013,Galletti2014,Kurter2014}. Specifically, we produced SQUIDs (Superconducting QUantum Interference Devices) on strained bulk HgTe, a 3D topological insulator, and studied their response to a DC magnetic field. In order to look for possible $p$-wave contributions, we investigated the symmetry of the order parameter by comparing two different geometries. One expects $p$- or $d$-wave symmetries to yield phase shifts in the magnetic response of the SQUID, depending on the orientation of the Josephson junctions contained in the circuit with respect to the orientations of the anisotropic superconductor order parameter \cite{Sigrist1995,Tsuei2000, Nelson2004}. The results show an absence of any phase shift in the SQUID response and of any clear deviation of the behavior from that of a conventional SQUID.

\section{Building blocks of the SQUIDs}
\label{section:JJ}
In this section, we succinctly describe the building blocks of the SQUIDs, namely HgTe as a 3D topological insulator, and Josephson junctions fabricated on this material. 

First, as described in \cite{Bruene2011}, strained layers of HgTe (76 nm thick) grown on CdTe substrates provide a 3D topological insulator with negligible bulk conductance when the Fermi level lies within its band gap. To characterize the transport properties of the material, Hall bars were defined by photo-lithography and Ar etching. For the layers on which SQUIDs were built, we extracted from the longitudinal and Hall resistances an electron density $n_e\simeq\SI{3.4e11}{\per\square\centi\meter}$ and a mobility $\mu\simeq \SI{5.3e4}{\square\centi\meter\per\volt\per\second}$. These values reflect the high quality of this material. Moreover, a series of quantization steps are visible in the Hall resistance at high magnetic fields, indicating that transport is indeed governed by two-dimensional surface states.

\begin{figure}[h!]
\centering\includegraphics[width=0.5\textwidth]{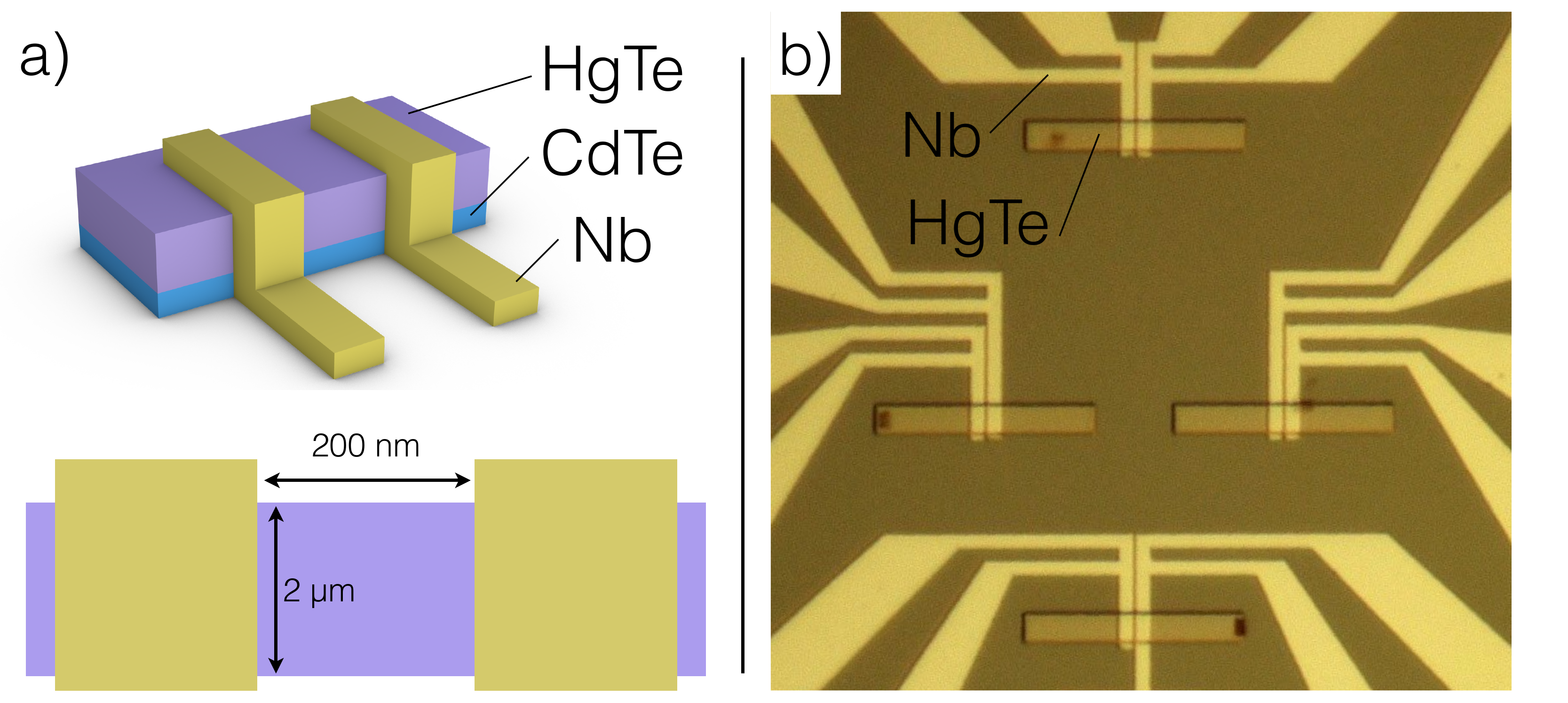}
\caption{ (a) Schematics of the lateral Josephson junctions. The strained growth of HgTe on CdTe provides a robust 3D topological insulator with negligible bulk conductance \cite{Bruene2011}. Nb contacts are then sputtered to create a lateral Josephson junction. The dimensions of the junctions presented in this article are $l_J=\SI{200}{\nano\meter}$ and $w_J=\SI{2}{\micro\meter}$. (b) Microscope image of the device presented in this section. The rectangular mesas of HgTe are visible as well as the Nb contacts, with two terminals on each contact allowing for 4-point measurements.} \label{fig:JJGeom}
\end{figure}

Lateral Josephson junctions were produced onto the surface of the HgTe \cite{Maier2012,Oostinga2013}: a mesa structure made of $\SI{2}{\micro\meter}$-wide stripes of HgTe is first realized by Ar etching. Nb contacts (68 nm thick) are added by sputtering. The geometry is described in Fig.\ref{fig:JJGeom}. The width of the junction is defined by the HgTe stripe ($w_J=\SI{2}{\micro\meter}$), while the most studied device had a distance $l_J=\SI{200}{\nano\meter}$ between the contacts.

\begin{figure}[h!]
\centering\includegraphics[width=0.5\textwidth]{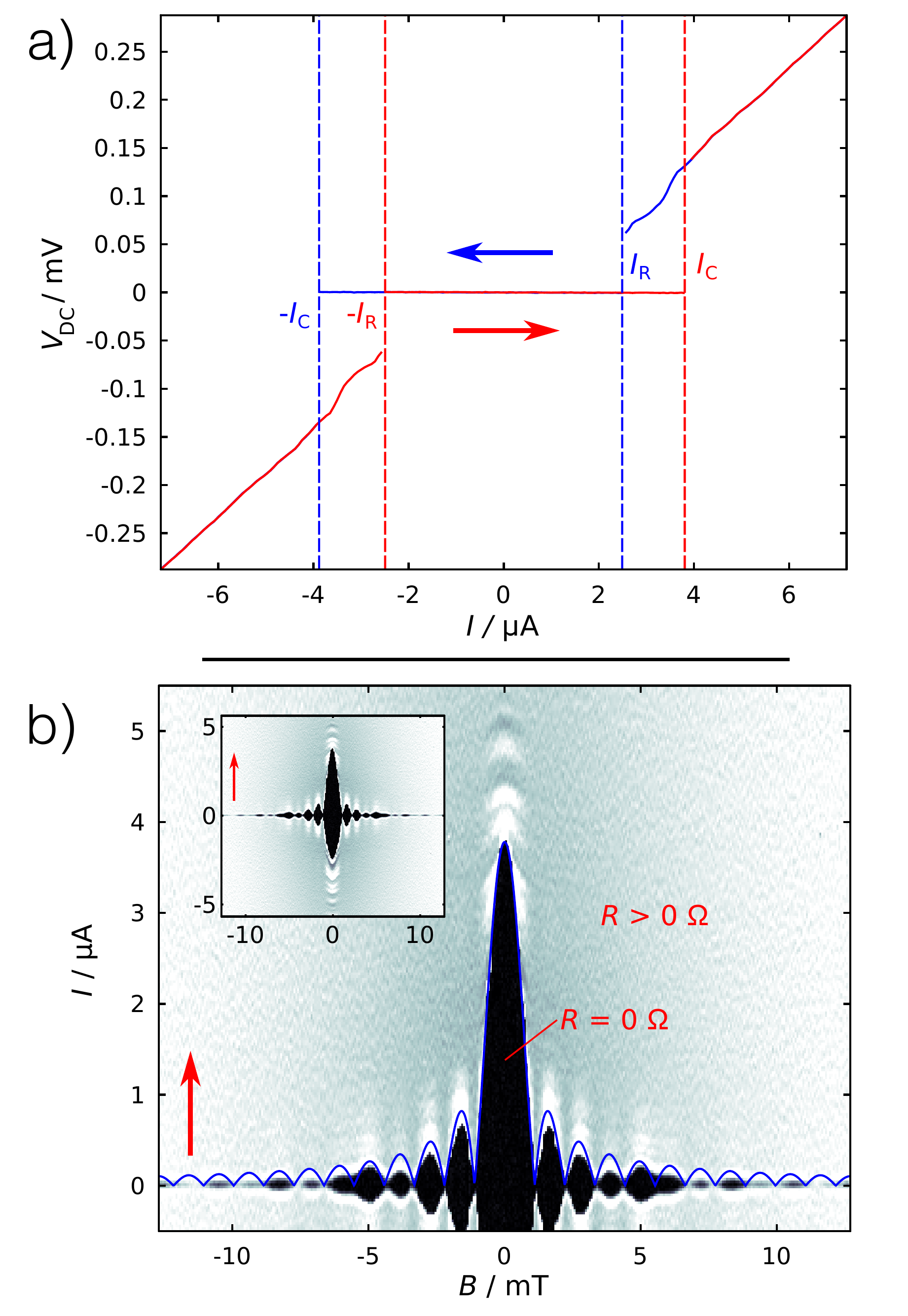}
\caption{ (a) I-V curves of one of the junctions. In blue and red lines, two curves show the results obtained for the two different sweep directions (indicated by the arrows). The junction presents a strong hysteretic character, with a critical current of $I_C=\SI{3.9}{\micro\ampere}$ and a retrapping current $I_R\simeq\SI{2.5}{\micro\ampere}$. (b) Magnetic field and current dependence of the same junction. The differential resistance $dV/dI$ across the junction is plotted in greyscale. The superconducting region is visible in black, with a resistance $R=0\ \Omega$, while the non-superconducting regions appear as grey areas, and exhibit $R>0\ \Omega$. The expected Fraunhofer pattern is clearly visible and a fit using Eq.\ref{eq:JJ} is shown as the blue line. The period of the pattern (1.1 mT) is in good agreement with a length of $l_J=\SI{200}{\nano\meter}$ if one takes into account the penetration depth of the magnetic field in the Nb contacts, $\lambda\simeq350$ nm. The red arrow symbolizes the sweep direction of the current. Due to hysteresis, the complete Fraunhofer pettern (shown in inset) is not symmetric with respect to $I=0$.} \label{fig:JJMeas}
\end{figure}

A typical I-V curve (measured at $T\simeq \SI{25}{\milli\kelvin}$) is shown in Fig.\ref{fig:JJMeas}. It can be described by two different regimes \cite{Oostinga2013}. First, when the bias is lower than a critical value $I_C\simeq\SI{3.9}{\micro\ampere}$, a dissipationless supercurrent occurs, and the junction exhibits zero resistance. For bias currents larger than $I_C$, a dissipating current flows through the junction, characterized by a normal state resistance of $R_n\simeq50\ \Omega$. Hysteretic behavior of the switching between these two regions is also observed. When sweeping back form large positive bias towards zero bias, one can identify the retrapping currents $I_R\simeq \SI{2.5}{\micro\ampere}$. From base temperature to 800 mK, we observed that the critical current decreases down to the value of $I_r$ which stays roughly constant in this range, so that hysteresis is typically visible up to 800 mK. Underdamping of the junction due to large self-capacitance \cite{McCumber1968,Tinkham2004} can be ruled out as the the geometrical capacitance $C_g$ of the junction is small. Thus the dimensionless McCumber parameter $\beta=\frac{2eI_CR_n^2C_g}{\hbar}\ll1$ tends to demonstrate that the junction is overdamped and no hysteresis is expected in this case. A plausible alternative is a self-heating effect\cite{Klapwijk1974,Courtois2008}.

More information can be drawn from differential resistance measurements (not shown here, see \cite{Oostinga2013}): some oscillations in the differential resistance reveal the presence of a Josephson supercurrent beyond the critical current $I_C$. Moreover, at large current bias, an excess current is measured ($I_{exc}\simeq\SI{4.5}{\micro\ampere}$), confirming the presence of Andreev bound states in the junctions.

Finally we explored the response of our junctions under a magnetic field oriented orthogonal to the sample plane. Under such conditions, the magnetic field induces a spatial modulation of the current distribution in the junction, leading to an interference pattern of the total current known as the Fraunhofer pattern. For a uniform junction, the critical current follows:
\begin{equation}
I_C(B)=I_C(0)\Big|\frac{\sin\pi\Phi_B/\Phi_0}{\pi\Phi_B/\Phi_0}\Big|\label{eq:JJ}
\end{equation}
where $\Phi_0=\frac{h}{2e}$ is the flux quantum, and $\Phi_B=A_{J}B$ is the flux threaded by the magnetic field $B$ through the surface $A_{J}$ of the junction. In Fig.\ref{fig:JJMeas}a, we present a two-dimensional grayscale plot of the differential resistance $dV/dI$ vs. the magnetic field $B$. The superconducting regions are seen in black while the non-superconducting areas appear in grey. The Fraunhofer pattern is clearly observable, with a periodicity of 1.1 mT, and a fit based on Eq.\ref{eq:JJ} is shown as the blue line. Our experimental data exhibit small sample-dependent deviations from the theoretical formula, indicating that the supercurrent is not fully uniform in the junction. From the periodicity, and the calculation of the area $A_J$ (taking into account the penetration length $\lambda=350$ nm of the B-field in the Nb contacts), we concluded that this corresponds to a $2\pi$-periodic critical current through the surface states. No $4\pi$-periodic modulation of the pattern is observed. This may be readily explained by the small influence of the unique $4\pi$-periodic topological mode with respect to the $2\pi$ ones, whose number is estimated between 50 to 200 (from $n_e$ and $w_J$) \cite{Tkachov2013}.

Thus, the supercurrent that developed in our HgTe-based Josephson junction does not exhibit any clear signature of topological superconductivity. However the technology developed for the fabrication of Josephson junctions also allows for a controlled production and measurement of SQUIDs that offer a much greater phase sensitivity. We detail the results obtained on these devices in the following section.

\section{Measurements of SQUIDs}

\begin{figure}[h!]
\centering\includegraphics[width=0.5\textwidth]{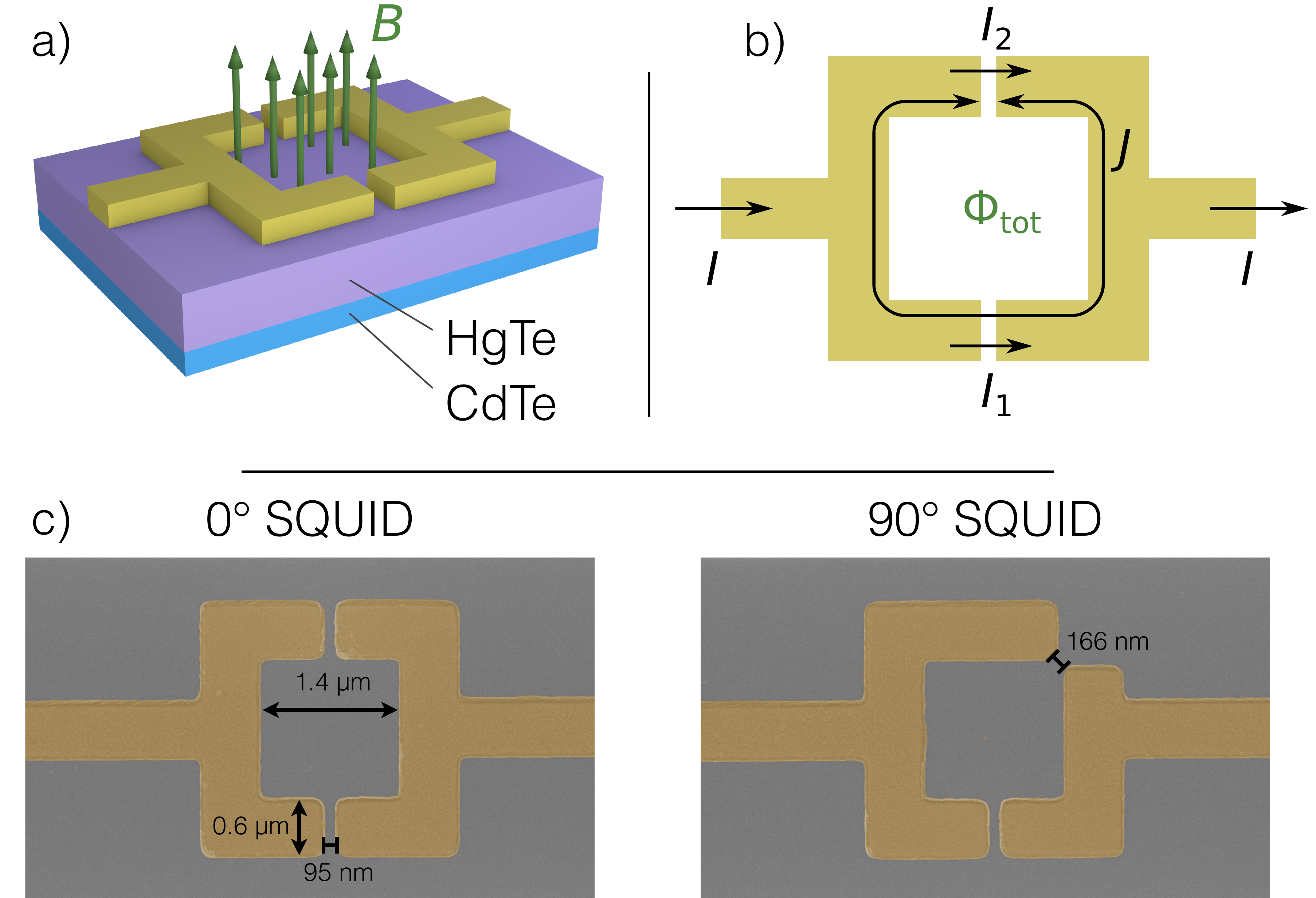}
\caption{ (a) Artist view of a SQUID device: the strained HgTe layer is grown on a CdTe substrate. A Nb SQUID (in yellow) is sputtered on the surface. A magnetic field $B$, symbolized by green arrows, is applied perpendicular to the HgTe surface to generate a flux $\Phi$ through the SQUID surface. (b) Schematic view of the SQUID: the bias current $I$ splits into two currents $I_1$ and $I_2$, flowing through the two Josephson junctions. (c) Colored SEM pictures of the two different SQUID geometries. Both geometries are based on a square of \SI{1.4}{\micro\meter} of inner dimensions with 600 nm wide Nb stripes. On the left, the $0^\circ$ SQUID comprises two straight Josephson junctions, with a length of 95 nm, while the $90^\circ$ SQUID is made of one straight junction and one $90^\circ$ junction with a minimal length of 166 nm.} \label{fig:SQUIDGeom}
\end{figure}

Using the above described lithographic processes, we defined SQUIDs comprising two weak links on a bare strained HgTe surface (Fig.\ref{fig:SQUIDGeom}a). The weak links are similar to the previously studied Josephson junctions, but two different geometries were explored. In the first geometry (referred to as the $0^\circ$-SQUID), the two junctions face each other on opposite sides of a square SQUID. The inner dimension of the square is $l_S=\SI{1.4}{\micro\meter}$. The junction has a width $w_J=600$ nm and a length $l_J$ of about 95 nm. All dimensions are indicated in Fig.\ref{fig:SQUIDGeom}. In the second geometry (referred to as the $90^\circ$-SQUID), the overall dimension of the square SQUID is maintained, but one of the junctions is placed in a corner, thus creating a $90^\circ$ angle junction, with a minimal (diagonal) distance of 166 nm.

Now let us briefly recall how a SQUID responds to an external perpendicular magnetic field $B$. As the phase of the macroscopic BCS wave function describing the superconductor must be single valued, the total magnetic flux $\Phi_{B}=A_SB$ (where $A_S$ is the inner area) through the SQUID is always such that $\Phi_{B}=n\Phi_0, n\in\mathbb{Z}$. The phase difference $\gamma$ between both superconducting regions should then follow the quantization condition $\gamma=2\pi\frac{\Phi_{B}}{\Phi_0}\rm\ mod\ 2\pi$. For two junctions with different critical currents $I_{C,i},i=1,2$, it is straightforward to obtain the critical current for the SQUID : 
\begin{equation}
I_{C,S}(B)=\sqrt{\big(I_{C,1}-I_{C,2}\big)^2+4I_{C,1}I_{C,2}\cos\frac{\pi\Phi_B}{\Phi_0}}.
\label{eq:SQUID}
\end{equation}
$I_{C,S}$ exhibits oscillations, with maxima for $\Phi_{B}=n\Phi_0$ and minima for $\Phi_{B}=(n+1/2)\Phi_0,n\in\mathbb{Z}$. The contrast of the oscillations is maximal for $I_{C,1}=I_{C,2}$ and will be strongly reduced for $I_{C,1}<I_{C,2}$ as expected for the $90^\circ$ SQUID. Moreover, the magnetic field affects the critical current of each junction due to their finite area, in agreement with the measurements of section \ref{section:JJ}. This causes a decay of the oscillations for larger fields (as the area of the ring is much wider than that of the junction, $A_S\gg A_{J}$). Note that the self-inductance of the SQUID (responsible for a circular current $J$, see Fig.\ref{fig:SQUIDGeom}) is expected to be negligible in our device and thus ignored in this derivation. 

\begin{figure}[h!]
\centering\includegraphics[width=0.5\textwidth]{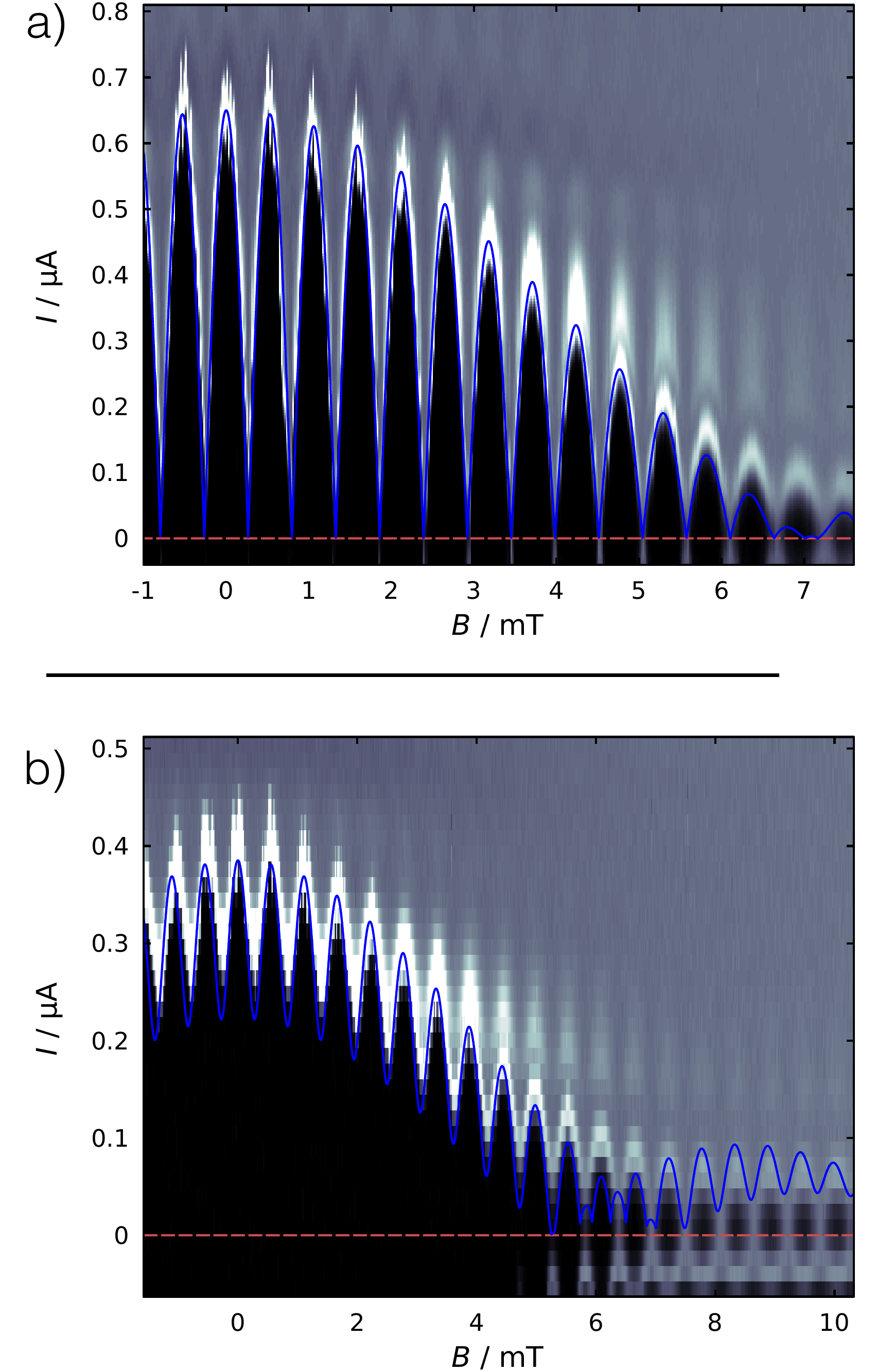}
\caption{ Magnetic field and current dependence of symmetric and asymmetric SQUIDs. The differential resistance $dV/dI$ of the junction is represented in greyscale. The black regions correspond to the superconducting regions (with $R\simeq 0\ \Omega$) while white and grey areas are finite resistance regions.  Fits of the critical current $I_{C,S}$ are obtained using Eq.\ref{eq:SQUID}. (a) For the $0^\circ$ SQUID, the critical currents of the two junctions are identical $I_{C,1}= I_{C,2}=\SI{0.33}{\micro\ampere}$ and the visibility of the oscillations is excellent. The period of the oscillations is 0.53 mT. (b) For the $90^\circ$ SQUID, the two critical currents are very different as expected for this asymmetric geometry: $I_{C,1}= \SI{0.081}{\micro\ampere}$, $I_{C,2}=\SI{0.33}{\micro\ampere}$ and the visibility of the oscillations is reduced. The period obtained is 0.56 mT. No measurable phase shift is observed between both devices.} \label{fig:SQUIDField}
\end{figure}

The experimental results are presented in Fig.\ref{fig:SQUIDField}. The differential resistance $dV/dI$ across the SQUID is shown in greyscale, as a function of the bias current $I$ and the magnetic field $B$. As previously, the dark regions correspond to the superconducting regions in which the SQUID resistance is $R=0\ \Omega$ while the white and grey areas correspond to a finite resistance state. This allows us to evaluate the critical current for each value of $B$. A fit relying on Eq.\ref{eq:SQUID} is shown as the blue line, with very good agreement to the data. In the case of the symmetric $0^\circ$ device (Fig.\ref{fig:SQUIDField}a), we obtain $I_{C,1}= I_{C,2}=\SI{0.33}{\micro\ampere}$, and very strong oscillations. On the other hand, the $90^\circ$ device (Fig.\ref{fig:SQUIDField}b) exhibits as expected very different critical currents: $I_{C,1}=\SI{0.081}{\micro\ampere}$, $I_{C,2}=\SI{0.33}{\micro\ampere}$, which dramatically reduces the contrast of the oscillations.
The periods of the $B$-field oscillations are similar in both cases, (0.53 mT and 0.56 mT for the $0/90^\circ$ SQUIDs respectively), which corresponds to an effective length of $l_S=\sqrt{A_S}\simeq\SI{1.97}{\micro\meter}$, in reasonable agreement with the dimensions of the SQUID if one takes into account the penetration length $\lambda$ on all sides of the square. 
We detect no phase shift between both geometries: assuming that the central lobe of the symmetric SQUID gives the reference zero field, no measurable difference is observed on the asymmetric one, within the accuracy of our measurements. The latter is dominated by the general drift of the magnetic field in the system: with an upper bound of \SI{2.8}{\micro\tesla\per\hour}, and a measurement time of approximately 48 h, the global error is less than 0.15 mT. The absence of any measurable phase difference between the two geometries thus tends to rule out pure $p$- or $d$-wave superconductivity, but is in agreement with the widely accepted model of proximity-induced $s$- and $p$- superconductivity in a topological insulator, for which the quasi-particle spectrum is isotropic \cite{Fu2008,Tkachov2013}.

\section{Conclusions}

We have fabricated SQUIDs in two geometries on HgTe and investigated the response of such devices to a perpendicular magnetic field. In order to investigate the symmetry of the superconducting order parameter, a $90^\circ$ SQUID is compared to a conventional $0^\circ$ SQUID. While a strong $p$-type pairing would be expected to give rise to a phase shift in the magnetic response of the $90^\circ$ device, we detect no such shift between both geometries. Thus, a pure $p$-wave symmetry can be excluded.

More generally, despite their phase sensitivity, Josephson junctions and SQUIDs on three-dimensional topological insulators (HgTe or BiSe-BiTe compounds) have not shown any clear signature of topological superconductivity \cite{Veldhorst2012,Galletti2014,Kurter2014}. Consequently, efforts should be made to reduce the number of channels participating in transport, and other methods should be considered such as the proposed measurements of the ac Josephson effect (Shapiro steps) \cite{Rokhinson2012, Dominguez2012}, finite-frequency noise \cite{Houzet2013} or S-TI-N junctions \cite{Alicea2012}.

\section*{Acknowledgements}

We gratefully acknowledge P. Sch\"uffelgen and C. Thurn for assistance during the measurements. This work was supported by the DARPA MESO Program through contract number N66001-11-1-4105, by the German Research Foundation (DFG Schwerpunkt 1666 'Topological Insulators', the DFG-JST joint research project 'Topological Electronics' and the Leibniz Program) and the EU ERC-AG program (project 3-TOP). E.B. and L.W.M. thank the Alexander von Humboldt Foundation for support through a Humboldt Research Fellowship. 

\section*{References}

\bibliographystyle{unsrt}
\bibliography{BibNobel.bib}

\end{document}